\documentclass[aps,prl,twocolumn,superscriptaddress,showpacs]{revtex4}

\usepackage{epsfig}
\usepackage{amsmath}

\begin{document}

\title{Universal Periods in Quantum Hall Droplets}

\author{Gregory A. Fiete}
\affiliation{Department of Physics, California Institute of Technology, MC 114-36, Pasadena, California 91125, USA}
\author{Gil Refael}
\affiliation{Department of Physics, California Institute of Technology, MC 114-36, Pasadena, California 91125, USA}
\author{Matthew P. A. Fisher}
\affiliation{Department of Physics, California Institute of Technology, MC 114-36, Pasadena, California 91125, USA}
\affiliation{Kavli Institute for Theoretical Physics, University of California, Santa Barbara, California 93106, USA}

\date{\today}

\begin{abstract}

Using the hierarchy picture of the fractional quantum Hall effect, we study the the ground state periodicity of a finite size quantum Hall droplet in a quantum Hall fluid of a different filling factor. The droplet edge charge is periodically modulated with flux through the droplet and will lead to a periodic variation in the conductance of a nearby point contact, such as occurs in some quantum Hall interferometers.  Our model is consistent with experiment and predicts that superperiods can be observed in geometries where no interfering trajectories occur. The model may also provide an experimentally feasible method of detecting elusive neutral modes and otherwise obtaining information about the microscopic edge structure in fractional quantum Hall states.

 \end{abstract}



\maketitle

With the recent surge of interest in quantum
computing\cite{DasSarma:pt06}, quantum Hall systems\cite{DasSarma}
have received renewed attention due to their potential use in topologically
protected qubits.   In particular, the $\nu=5/2$ and
$\nu=12/5$ quantum-Hall states are believed to support non-Abelian
excitations\cite{Moore:npb91} which are crucial
ingredients for topological quantum computation\cite{DasSarma:pt06}.  Here we will only focus on the Abelian quantum Hall states, but we will make use of their topological properties to reveal universal periodicities (as a function of magnetic flux through the droplet) in the ground state energy and edge properties of a quantum Hall droplet inside a surrounding Hall fluid of a different filling factor. The universal periodicities in the ground state properties can generically be used to
probe the quantum Hall edge states in equilibrium settings.

\begin{figure}[b]
\includegraphics[width=.65\linewidth,clip=]{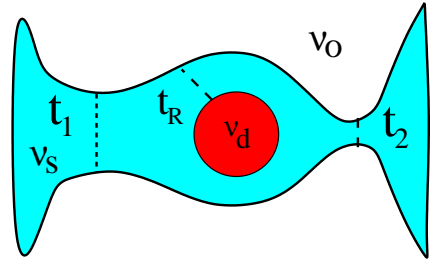}
 \caption{\label{fig:schematic} (color online) Schematic of our setup.
 A quantum Hall droplet of filling factor $\nu_d$ is surrounded by a
 Hall fluid with $\nu_s \neq \nu_d$, itself surrounded by an outer
 fluid with $\nu_o$. Tunneling between the $\nu_s$  fluid edges occurs at two point
 contacts with amplitudes $t_1, t_2$. Tunneling (with amplitude $t_R$) also occurs between
 the droplet edge and the surrounding fluid edge, which acts as a
 reservoir. Periodic charging of the droplet edge
 with flux will cause periodic modulations of the tunneling amplitudes
 $t_1,t_2$ implying conductance oscillations of the same period at the
 point contacts. The period will depend on $\nu_s,\,\nu_d$ and $\nu_o$
 with periods greater than a flux quantum possible.  The system could
 be contacted in the way desribed in Ref.\cite{Camino:prl05}. }
 \end{figure}

Our work is motivated in part by a series of beautiful experiments
done on quantum Hall interferometers where superperiods and fractional statistics have purportedly been
observed\cite{Camino:prl05}.   Several
theoretical studies have already addressed these
experiments\cite{Kim:prl06} but a complete
picture, particularly in the fractional quantum Hall regime, is still
lacking.  In this Letter we study the universal properties of a finite
size quantum Hall droplet inside a quantum Hall fluid of a different
filling factor (Fig. \ref{fig:schematic}). For most geometries
and droplet filling factors we find that the ground state energy of
the system has a periodicity with magnetic flux through the inner droplet
that is determined only by the two filling fractions in the limit that
the charging energy of the surrounding fluid edge tends to zero.  However, edges such as that of the $\nu=2/3$ state (which have counter propagating modes and disorder influenced excitations\cite{Kane:prl94}) require a special degree of consideration, as we discuss below.

Consider a droplet of filling factor $\nu_d$ surrounded by a fluid of
filling factor $\nu_s$ which itself may be inside an outer fluid of
filling $\nu_o$ (Fig. \ref{fig:schematic}). As magnetic flux is
adiabatically threaded through the inner droplet, its
ground-state energy and its radius oscillate with a universal
periodicity.  This periodicity reveals important
information about the microscopic structure of the droplet edge itself, thus
providing a mechanism by which theoretical edge models can be directly
tested experimentally. We concentrate on two important special cases: (i)  $\nu_d=2/5,\,\nu_s=1/3,\,\nu_o=0$ and (ii)
$\nu_d=2/3,\,\nu_s=0,\,\nu_o=1$.  General filling fractions with
Abelian statistics follow one of the two cases above. While
disorder does not play a central role in the physics of the edge in case
(i), in case (ii) disorder determines the nature of the edge
excitations \cite{Kane:prl94}, by causing
counter-propogating modes of the droplet edge to ``recombine" into charged and neutral
modes. The precise way in which the radius changes with flux can directly probe whether this recombination occurs, or
whether the edge structure is that of the clean system, as described in
Ref.\cite{MacDonald:prl90}.  Therefore, the flux dependence of the conductance can reveal the presence or absence of elusive neutral modes, which to the best of our knowledge have not yet been experimentally detected.

The expansion and contraction of the inner droplet results from a
charging and discharging of its edge. In the proposed setup, this affects the conductance of
a quantum point contact near the droplet due to Coloumb
interaction: the changing electric potential near the point
contact affects the distance between the two edges that the point contact connects.
Such an interaction-modulated conductance was used to detect charge states in
a double quantum dot system and to coherently
manipulate spin\cite{DiCarlo:prl04}.  It might also be relevant for certain geometries of quantum Hall
interferometers, such as those of
Ref.\cite{Camino:prl05} and may be the cause of the superperiods observed there, rather than
interference. In fact, superperiods would also result from potential modulations at
a {\em single} point contact via the physics outlined above. Whether
the origin of observed superperiods in interferometers with two point contacts
are the result of interaction-modulated tunneling, or of true
interference, can be determined in experiment by suppressing tunneling
at one of the point contacts ($t_1$ in Fig.~\ref{fig:schematic}, for example). Interference effects would disappear, but interaction modulated tunneling effects would still produce
oscillations.  A similar suggestion was also made be Rosenow and Halperin in Ref.\cite{Kim:prl06}.

The ground state periodicity for a quantum Hall droplet has been
discussed before\cite{Jain:prl93} using bulk fluid
descriptions. Here we use an edge state description \cite{Wen:ijmp92} to emphasize the physics
that depends on the nature of the edge modes themselves.  As we are interested in the
universal properties of such a droplet in a surrounding Hall fluid, we use the theory describing the universal
aspects of the fractional quantum Hall state\cite{Wen:ijmp92,Zhang:prl89}. The edge modes constitute a minimal model\cite{reconst_comment} described
by the $K-$matrix formulation of Wen\cite{Wen:ijmp92}.  In
this formulation, the action is
\begin{equation}
S_{\rm edge}=\int \frac{dt dx}{4\pi}[K_{ij}\partial_t \phi_i \partial_x
  \phi_j -V_{ij}\partial_x \phi_i \partial_x \phi_j-2t_iA \partial_x \phi_i],
\label{Sedge}
\end{equation}
where $K$ is a matrix determined by a choice of basis
denoted by $t$ which determines the coupling to the vector potential $A$; the filling is $\nu=t^\dagger
  K^{-1} t$.  The dimension of $K$ reflects the filling $\nu$ of the
quantum Hall state, and equals the number of independent edge modes. $V$ is a {\em non-universal} positive
definite matrix determined by edge mode interactions and the confining
potential, and $\phi_i(x,t)$ are bosonic fields parameterizing the edge
modes. 

The topological content of the quantum Hall state is encoded in
$K$. For a droplet inside a surrounding fluid, the interface
$K-$matrix is \cite{Kao:prl99}
\begin{equation}\label{eq:K}
K=\left(\begin{array}{cc}
K_d & 0\\
0 & -K_s
\end{array}\right),
\end{equation}  
where $K_d$ and $K_s$ describe the droplet and sorrounding liquid
respectively. If there exists an integer valued vector $m$ such that $m^\dagger
K^{-1} m=0$ and $t^\dagger K^{-1} m=0$, then $K$ is topologically
unstable\cite{Haldane:prl95} and may
reconstruct by some edge modes ``gapping'' each other out, thus
reducing the number of edge modes. If the droplet Hall state is a descendant of the
surrounding state, this is always possible, and akin to low level
composite fermion Landau levels connecting adiabatically across the interface.
  The topological
stability of the interface edge also depends on the non-universal $V$; here we
will assume instability, since it occurs for a wide and realistic range of $V$.

A crucial component for our setup is the finite size of the
droplet. This implies a level quantization, which can be inferred
using gauge invariance and quantized Hall
conductance. Consider an edge described by the field
$\phi$, which is a linear combination of the $\phi_i$ that diagonalizes the matrices $K$ and $V$, and obeys $[\phi(x),\partial_x\phi(x')]=q\delta(x-x')$. The
operator $\exp(i\phi)$ thus creates an edge excitation of charge
$q$ and upon flux $h/e$ insertion at a point within the inner droplet, the
creation operator must become:
$\exp(i\phi)\rightarrow \exp(i\phi+2\pi i q \frac{x}{L}),$
where $L$ is the length of the edge. But from gauge invariance, the spectrum of the edge must remain
unchanged.  The charge in each of the orbitals must then
be $q$ to obtain quantized Hall conductance, $\sigma_{xy}=q
e^2/h$. Thus, a finite edge loop can be described as a chiral
Luttinger liquid which consists of discrete orbitals, each containing
charge $q$. 
 
Let us now treat the case of a $\nu_d=2/5$ droplet in a
$\nu_s=1/3$ and $\nu_o=0$ surrounding. For filling factor
$\nu=n/(np+1)$, the $K-$matrix in the symmetric basis is an
$n-$dimensional matrix\cite{Kane:prl94}, $K_{ij}=\delta_{ij}+p$. Since
the 2/5 state is the daughter of the 1/3 state, the $K$ matrix given by Eq.\eqref{eq:K} is
indeed unstable, and the resulting recombined edge is identical to
that of a $\nu=1/15$ Laughlin state.  The $1/15$ effective edge within
a $1/3$ edge, leads to a $5 \Phi_0$ periodicity of ground state
properties of the droplet with magnetic
flux through it, a result in agreement with a bulk
description\cite{Jain:prl93} and experiment\cite{Camino:prl05}. 

To see this, consider the gapped droplet state. Upon an adiabatic
$\Phi_0=h/e$ flux insertion at a point in
the $\nu_d=2/5$ droplet, a net charge of $2e/5$ is localized at the
flux. This charge is sucked from the two edges: an
$e/15$ orbital is vacated in the $2/5-1/3$ edge, and an additional
$e/3$ orbital is vacated in the $1/3-0$ edge. Confirming our assertion
above as to the edge structure, indeed $1/3+1/15=2/5$. Smearing the flux uniformly over the droplet yields the same result. Repeating the adiabatic flux insertion will progressively charge the droplet edge in units of $-e/15$
and the surrounding fluid edge in units of $-e/3$.  Additional flux
outside the droplet may create additional excitations of charge $-e/3$
on the outer edge, but it will not influence the edge charge of the
droplet. Through the ubiquitous presence of disorder is quantum Hall systems, it is possible for quasi-particles to
tunnel\cite{Auerbach:prl98} between the $1/3-0$ edge
and the $2/5-1/3$ edge, and relax the energy of the
system.  The allowed charges are determined by
$K_s$\cite{Wen:ijmp92}, and the most relevant operator in the present case
is indeed $e^{(-i\phi_s+5i\phi_d)}$, which tunnels charge $e/3$.

\begin{figure}[t]
 \includegraphics[width=.7\linewidth,clip=]{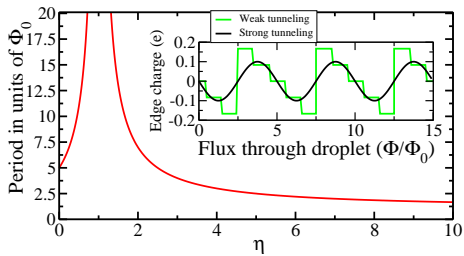}%
 \caption{\label{fig:periods} (color online) Periodicity of ground state
 structure vs. flux through a $\nu_d=2/5$ droplet in a $\nu_s=1/3$
 surrounding Hall fluid. $\eta=\frac{E_c^s}{5E_c^d}$ is
 described in the text. Note the period is non-universal unless one
 edge is very long compared to the other. When $\eta \to 0$
 the periodicity is independent of the flux through the surrounding
 Hall fluid and is equal to a universal value, $5\Phi_0$.  Inset:
 Droplet edge charge in the universal limit vs. flux for
 weak and strong tunneling $t_R$.}
 \end{figure}
 
Assuming both edges are initially neutral, denote the number of 
additional filled edge states by $n_d$ and $n_s$, for the droplet and
surrounding fluid respectively.  The energy of the charged edges is 
$E_{d/s}=\frac{E_c^{d/s}}{2}n_{d/s}^2$ for the droplet/surrounding fluid. The energies
$E_c^d,E_c^s$ depend on the the edge
velocities and capacitances, and are inversely proportional to the
length of the edges.  The total charge on the
edges is $Q=-\frac{e}{15}n_d-\frac{e}{3}n_s$.  The two distinct edge
excitations have chemical potentials
determined by $\mu \equiv \frac{\partial E}{\partial n}$ which gives
$\mu_{d/s}=E_c^{d/s} n_{d/s}$. When the two edges are in
equilibrium, $\mu_d \delta n_d + \mu_s \delta n_s=0$, and from charge
conservation, $\delta n_d=-5\delta n_s$. Thus
$5\mu_d=\mu_s$ (i.e. the two edges have the same {\it voltage}). This
also gives $E_c^s n_s=5 E_c^d n_d$ at edge equilibrium. Now, the edge
occupations $n_d$ and $n_s$ depend on the flux threaded.
Assuming all of the flux is through the droplet, we have $6
\frac{\Phi}{\Phi_0}=n_d + 5 n_s=(\eta+5)n_s$ where $\eta\equiv
\frac{E_c^s}{5E_c^d}$.  Solving this for $n_s$ gives:
$n_s=\mbox{round}\left[\frac{\Phi}{\Phi_0}
+\left(\frac{1-\eta}{5+\eta}\right)\frac{\Phi}{\Phi_0}\right].$

The first term indicates that every flux insertion raises $n_s$ by
 one. The second describes $e/3$ charge transfer between the two edges.  In the limit $\eta \to 0$ (occuring when the length
of the surrounding fluid edge is long compared to the droplet edge),
every $5\Phi_0$ added  increases $n_s$ by one extra state due to the tunneling of an $e/3$ charge. This
 happens when the rounding function changes from rounding down to
 up.  The one $e/3$ charge annihilates 5 $-e/15$ charges and returns the droplet edge
to its initial state.  In the
opposite limit, $\eta \to \infty$, every $\nu=1/3$ orbital vacated due
 to $\Phi_0$ insertion immediately gets filled via charge transfer
 {\em from} the droplet edge. For
general values of $\eta$ (i.e., the ratio of  edge ``charging
energies") the period is non-universal as shown in
Fig. \ref{fig:periods}.  Allowing flux insertion in both the droplet
and the surrounding fluid, changes the response of the $1/3$ edge,
 but in the limit  $\eta \to 0$, this will not affect the period with
 respect to droplet flux, as the ``rate limiting" step is
due to the finite compressiblity of the droplet edge.  Therefore, the
droplet $5\Phi_0$ flux period emerges as the universal droplet edge charging result
when $\eta \to 0$, independent of the area of the
  surrounding Hall fluid.  
  
\begin{figure}[h]
\includegraphics[width=.7\linewidth,clip=]{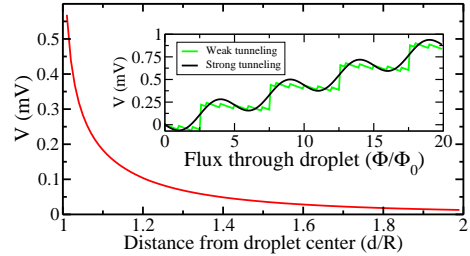}
\caption{\label{fig:potential} (color online) Potential created by a
  disk and ring of opposite charge equal to that of one electron.  The
  dielectric constant is assumed to be that for GaAs,
  $\epsilon=12$. Inset: Different form of the periodic voltage
  modulations depending on tunneling $t_R$ between the droplet edge
  and the surrounding fluid edge with $d/R=1.1$ for a $\nu_d=2/5$
  droplet in a $\nu_s=1/3$ surrounding fluid. Subtracting a smooth
  background will lead to oscillations like those in the inset of
  Fig.\ref{fig:periods}.}
 \end{figure}

A finite tunneling amplitude between the two edges foils the exact
quantization of the expectation value of the droplet
edge charge\cite{MatveevQdot}, as shown in the inset of
Fig.~\ref{fig:periods} in the limit $\eta \to 0$. 
With the smooth oscillation of edge charge, there is a flux dependent oscillation of the electrical potential that will affect the conductance of a
nearby point contact, as in Fig.\ref{fig:schematic}.  Modeling the droplet as
an outer ring of charge $Q_{\rm ring}$ and a uniformly charged
inner disk of net charge $Q_{\rm disk}$, the potential from a droplet
of radius $R$ at a distance $d>R$ from the center is $V=V_{\rm
  ring}+V_{\rm disk}$ where $V_{\rm ring}=\frac{Q_{\rm ring}}{\epsilon
  \pi}\left(\frac{K(-4Rd/(d-R)^2)}{d-R}+\frac{K(4Rd/(d+R)^2)}{d+R}\right)$ and $V_{\rm disk}=\frac{2Q_{\rm disk}}{\epsilon \pi R^2}\int_0^R r dr\left(\frac{K(-4rd/(d-r)^2)}{d-r}+\frac{K(4rd/(d+r)^2)}{d+r}\right)$.  Here $\epsilon$ is the dielectric constant and $K(x)$ is the complete elliptic integral of the first kind. The potential fluctuations are plotted in Fig.~\ref{fig:potential} and should be observable.  A metallic gate placed 100-200 nm above the droplet will reduce the potential modulations by no more than 30\%. 

Let us now focus on a case where the droplet
edge has counter propagating modes: $\nu_d=2/3,\,\nu_s=0,$ and $\nu_o=1$.
The $\nu_d=2/3$ to zero edge itself has
counter-propagating modes which leads to a disorder-dependent edge
structure. The clean $2/3$ edge has an outer $\nu=1$
mode and an inner (counter-propagating) $\nu=-1/3$
mode\cite{MacDonald:prl90,Zulicke:prb98}. But as
Ref.\cite{Kane:prl94} predicts, in the disorder-dominated phase, the effective low-energy degrees of freedom are a charge mode with
$q=2e/3$, that gives a quantized Hall conductance, and a counter-propagating 
neutral mode localized when $T\neq0$.  

\begin{figure}[h]
\includegraphics[width=.75\linewidth,clip=]{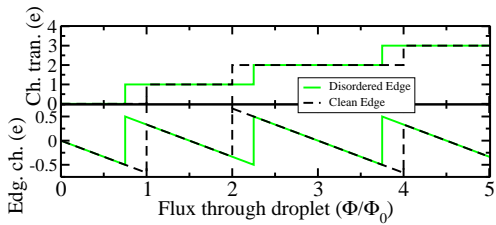}
\caption{\label{fig:23edge}(color online) Comparison of two different
  phases of the $\nu_d=2/3$ edge in the $\nu_s=0,\nu_o=1$. Top: Electron transfer vs. flux
  through the droplet. Bottom: Droplet edge charge vs. flux through the droplet. The microscopic edge structure determines
  the flux dependence of the charge transfer and electrical potential
  created at a nearby point contact, as in Fig.~\ref{fig:schematic}. }
 \end{figure}

The ``surrounded droplet" setup allows an equilibrium
verification of the charge/neutral recombination scenario.  When flux is threaded through
the $\nu_d=2/3$ droplet, the edge charge 
relaxes through {\it electron} tunneling across the vacuum. For
the disorder-dominated edge, the edge charge orbitals effectively consist of a single
$\nu=2/3$ mode as the neutral modes do not respond to flux. (This requires the droplet edge be long compared to the recombination length of the neutral/charge modes, which is sample dependent and finite even  at $T=0$.) Here, after every $3\Phi_0/2$ flux threading, the
edge loses one electron, and can then discharge by an electron tunneling from the
$\nu_0=1$ fluid. But for the clean phase, the flux
dependence of the edge charge is different.  When $\Phi_0$ is threaded
through the droplet the clean 2/3 edge accumulates a $-e$ charge on the
outer $\nu=1$ mode and a $e/3$ charge on the inner mode. In the limit
of large $\nu_o=1$ edge length, an
electron from the outer edge tunnels in to lower
the energy. This continues when a second $\Phi_0$ is threaded, but
when the third $\Phi_0$ is threaded, the edge instead relaxes to its
original state by
three $e/3$ inner mode excitations canceling one $-e$ outer mode
excitation. This sequence is shown in
Fig. \ref{fig:23edge}.  Fourier transforming the signal should
allow for a clear indentification of each case; one
charge mode on the droplet edge leads to one periodicity appearing, whereas the two independent charge modes of the clean edge should exhibit two periodicities. Other edges with
counter-propagating modes (such as $\nu_d=3/5$) are amenable to similar
considerations.

In this Letter we propose  the ``surrounded droplet" model
near a point contact to investigate universal properties of composite edges. The
periodic change of the droplet size with flux is measured by its
effect on the conductance on a nearby point contact. We
propose to use this effect to explore the nature of the $\nu=2/3$
edge, i.e., whether the edge recombines into a neutral and
charged modes. This is the first proposal that may be able
to do so in equilibrium. We assumed that the interior of
all Hall droplets are gapped, and that the only compressible areas are
at the bounderies, neglecting the possibility that Hall droplets may break down into incompressible
and compressible regions\cite{Chklovskii:prb92}. With sufficient disorder, all
quasiparticle states in the interior compressible regions are localized, keeping our analysis intact. A back-gate close to the
sample, however, will avoid this complication altogether, with the relevant
length scale being of order 200nm, i.e. comparable to the distance between the (in-plane) front gate in Ref.\cite{Chklovskii:prb92} and the outer edge of the hall droplet. This will impose a
rather uniform chemical potential on the electronic fluid, and hamper
the creation of compressible domains but should still allow sufficient potential modulation at the point contact to observe the predicted effects.

We thank Jim Eisenstein, Bernd Rosenow, and Xiao-Gang Wen for enlightening
discussions. Financial support from NSF Grants PHY05-51164 and
DMR05-29399 is gratefully acknowledged.  GAF was also supported by the
Lee A. DuBridge Foundation and MPAF by the Moore Foundation.


\begin{thebibliography}{19}
\expandafter\ifx\csname natexlab\endcsname\relax\def\natexlab#1{#1}\fi
\expandafter\ifx\csname bibnamefont\endcsname\relax
  \def\bibnamefont#1{#1}\fi
\expandafter\ifx\csname bibfnamefont\endcsname\relax
  \def\bibfnamefont#1{#1}\fi
\expandafter\ifx\csname citenamefont\endcsname\relax
  \def\citenamefont#1{#1}\fi
\expandafter\ifx\csname url\endcsname\relax
  \def\url#1{\texttt{#1}}\fi
\expandafter\ifx\csname urlprefix\endcsname\relax\def\urlprefix{URL }\fi
\providecommand{\bibinfo}[2]{#2}
\providecommand{\eprint}[2][]{\url{#2}}

\bibitem[{\citenamefont{{Das Sarma} et~al.}(2006)\citenamefont{{Das Sarma},
  Freedman, and Nayak}}]{DasSarma:pt06}
\bibinfo{author}{\bibfnamefont{S.}~\bibnamefont{{Das Sarma}}},
  \bibinfo{author}{\bibfnamefont{M.}~\bibnamefont{Freedman}}, \bibnamefont{and}
  \bibinfo{author}{\bibfnamefont{C.}~\bibnamefont{Nayak}},
  \bibinfo{journal}{Physics Today} \textbf{\bibinfo{volume}{59}},
  \bibinfo{pages}{32} (\bibinfo{year}{2006});
%
\bibinfo{author}{\bibfnamefont{A.}~\bibnamefont{Kitaev}},
  \bibinfo{journal}{Ann. Phys.} \textbf{\bibinfo{volume}{303}},
  \bibinfo{pages}{2} (\bibinfo{year}{2003}).
  
\bibitem[{Das()}]{DasSarma}
\bibinfo{note}{See {\it Perspectives in Quantum Hall Effects} (Wiley and Sons,
  Inc., New York, New York, 1997), Eds. S. Das Sarma and A. Pinczuk, and
  references therin.}

\bibitem[{\citenamefont{Moore and Read}(1991)}]{Moore:npb91}
\bibinfo{author}{\bibfnamefont{G.}~\bibnamefont{Moore}} \bibnamefont{and}
  \bibinfo{author}{\bibfnamefont{N.}~\bibnamefont{Read}},
  \bibinfo{journal}{Nucl. Phys. B} \textbf{\bibinfo{volume}{360}},
  \bibinfo{pages}{362} (\bibinfo{year}{1991});
%
\bibinfo{author}{\bibfnamefont{N.}~\bibnamefont{Read}} \bibnamefont{and}
  \bibinfo{author}{\bibfnamefont{E.}~\bibnamefont{Rezayi}},
  \bibinfo{journal}{Phys. Rev. B} \textbf{\bibinfo{volume}{59}},
  \bibinfo{pages}{8084} (\bibinfo{year}{1999}).


\bibitem[{\citenamefont{Camino et~al.}(2005{\natexlab{a}})\citenamefont{Camino,
  Zhou, and Goldman}}]{Camino:prl05}
\bibinfo{author}{\bibfnamefont{F.~E.} \bibnamefont{Camino {\it et al.}}},
 \bibinfo{journal}{Phys. Rev. Lett.} \textbf{\bibinfo{volume}{95}},
  \bibinfo{eid}{246802} (\bibinfo{year}{2005}{\natexlab{a}});
%
\bibinfo{author}{\bibfnamefont{F.~E.} \bibnamefont{Camino {\it et al.}}},
  \bibinfo{journal}{Phys. Rev. B} \textbf{\bibinfo{volume}{72}},
  \bibinfo{eid}{075342} (\bibinfo{year}{2005}{\natexlab{b}}).


\bibitem[{\citenamefont{Kim}(2006)}]{Kim:prl06}
\bibinfo{author}{\bibfnamefont{E.-A.} \bibnamefont{Kim}},
  \bibinfo{journal}{Phys. Rev. Lett.} \textbf{\bibinfo{volume}{97}},
  \bibinfo{eid}{216404} (\bibinfo{year}{2006});
%
\bibinfo{author}{\bibfnamefont{B.}~\bibnamefont{Rosenow}} \bibnamefont{and}
  \bibinfo{author}{\bibfnamefont{B.~I.} \bibnamefont{Halperin}},
  \bibinfo{journal}{{\it ibid}.} \textbf{\bibinfo{volume}{98}},
  \bibinfo{eid}{106801} (\bibinfo{year}{2007});
%
\bibinfo{author}{\bibfnamefont{J.~K.} \bibnamefont{Jain}} \bibnamefont{and}
  \bibinfo{author}{\bibfnamefont{C.}~\bibnamefont{Shi}},
  \bibinfo{journal}{{\it ibid}.} \textbf{\bibinfo{volume}{96}},
  \bibinfo{eid}{136802} (\bibinfo{year}{2006}).

\bibitem[{\citenamefont{Kane et~al.}(1994)\citenamefont{Kane, Fisher, and
  Polchinski}}]{Kane:prl94}
\bibinfo{author}{\bibfnamefont{C.~L.} \bibnamefont{Kane}},
  \bibinfo{author}{\bibfnamefont{M.~P.~A.} \bibnamefont{Fisher}},
  \bibnamefont{and}
  \bibinfo{author}{\bibfnamefont{J.}~\bibnamefont{Polchinski}},
  \bibinfo{journal}{Phys. Rev. Lett.} \textbf{\bibinfo{volume}{72}},
  \bibinfo{pages}{4129} (\bibinfo{year}{1994});
%
\bibinfo{author}{\bibfnamefont{C.~L.} \bibnamefont{Kane}} \bibnamefont{and}
  \bibinfo{author}{\bibfnamefont{M.~P.~A.} \bibnamefont{Fisher}},
  \bibinfo{journal}{Phys. Rev. B} \textbf{\bibinfo{volume}{51}},
  \bibinfo{pages}{13449} (\bibinfo{year}{1995}).

\bibitem[{\citenamefont{MacDonald}(1990)}]{MacDonald:prl90}
\bibinfo{author}{\bibfnamefont{A.~H.} \bibnamefont{MacDonald}},
  \bibinfo{journal}{Phys. Rev. Lett.} \textbf{\bibinfo{volume}{64}},
  \bibinfo{pages}{220} (\bibinfo{year}{1990});
%
\bibinfo{author}{\bibfnamefont{X.~G.} \bibnamefont{Wen}},
  \bibinfo{journal}{{\it ibid}.} \textbf{\bibinfo{volume}{64}},
  \bibinfo{pages}{2206} (\bibinfo{year}{1990}).

\bibitem[{\citenamefont{DiCarlo et~al.}(2004)\citenamefont{DiCarlo, Lynch,
  Johnson, Childress, Crockett, Marcus, Hanson, and Gossard}}]{DiCarlo:prl04}
\bibinfo{author}{\bibfnamefont{L.}~\bibnamefont{DiCarlo {\it et al}.}},
  \bibinfo{journal}{Phys. Rev. Lett.}
  \textbf{\bibinfo{volume}{92}}, \bibinfo{eid}{226801} (\bibinfo{year}{2004});
%
\bibinfo{author}{\bibfnamefont{J.}~\bibnamefont{Petta {\it et al.}}},
  \bibinfo{journal}{Science} \textbf{\bibinfo{volume}{309}},
  \bibinfo{pages}{2180} (\bibinfo{year}{2005}).

\bibitem[{\citenamefont{Jain et~al.}(1993)\citenamefont{Jain, Kivelson, and
  Thouless}}]{Jain:prl93}
\bibinfo{author}{\bibfnamefont{J.~K.} \bibnamefont{Jain}},
  \bibinfo{author}{\bibfnamefont{S.~A.} \bibnamefont{Kivelson}},
  \bibnamefont{and} \bibinfo{author}{\bibfnamefont{D.~J.}
  \bibnamefont{Thouless}}, \bibinfo{journal}{Phys. Rev. Lett.}
  \textbf{\bibinfo{volume}{71}}, \bibinfo{pages}{3003} (\bibinfo{year}{1993});
%
\bibinfo{author}{\bibfnamefont{A.~H.} \bibnamefont{MacDonald}}
  \bibnamefont{and} \bibinfo{author}{\bibfnamefont{M.~D.}
  \bibnamefont{Johnson}}, \bibinfo{journal}{{\it ibid}.}
  \textbf{\bibinfo{volume}{70}}, \bibinfo{pages}{3107} (\bibinfo{year}{1993}).

\bibitem[{\citenamefont{Wen}(1992)}]{Wen:ijmp92}
\bibinfo{author}{\bibfnamefont{X.~G.} \bibnamefont{Wen}},
  \bibinfo{journal}{Int. J. Mod. Phys.} \textbf{\bibinfo{volume}{B6}},
  \bibinfo{pages}{1711} (\bibinfo{year}{1992}).

\bibitem[{\citenamefont{Zhang et~al.}(1989)\citenamefont{Zhang, Hansson, and
  Kivelson}}]{Zhang:prl89}
\bibinfo{author}{\bibfnamefont{S.~C.} \bibnamefont{Zhang}},
  \bibinfo{author}{\bibfnamefont{T.~H.} \bibnamefont{Hansson}},
  \bibnamefont{and} \bibinfo{author}{\bibfnamefont{S.}~\bibnamefont{Kivelson}},
  \bibinfo{journal}{Phys. Rev. Lett.} \textbf{\bibinfo{volume}{62}},
  \bibinfo{pages}{82} (\bibinfo{year}{1989});
%
\bibinfo{author}{\bibfnamefont{D.-H.} \bibnamefont{Lee}} \bibnamefont{and}
  \bibinfo{author}{\bibfnamefont{M.~P.~A.} \bibnamefont{Fisher}},
  \bibinfo{journal}{{\it ibid}.} \textbf{\bibinfo{volume}{63}},
  \bibinfo{pages}{903} (\bibinfo{year}{1989});
%
\bibinfo{author}{\bibfnamefont{N.}~\bibnamefont{Read}}, \bibinfo{journal}{{\it ibid}.} \textbf{\bibinfo{volume}{65}}, \bibinfo{pages}{1502}
  (\bibinfo{year}{1990}).

\bibitem[{rec()}]{reconst_comment}
\bibinfo{note}{At finite temperatures edge reconstruction rapidly vanishes: Y.
  N. Joglekar {\it et al.}, Phys. Rev. B {\bf 68}, 035332
  (2003). Wen's formulation also assumes the edges are sufficiently sharp.}


\bibitem[{\citenamefont{Kao et~al.}(1999)\citenamefont{Kao, Chang, and
  Wen}}]{Kao:prl99}
\bibinfo{author}{\bibfnamefont{H.-c.} \bibnamefont{Kao {\it et al.}}},
  \bibinfo{journal}{Phys. Rev. Lett.} \textbf{\bibinfo{volume}{83}},
  \bibinfo{pages}{5563} (\bibinfo{year}{1999}).

\bibitem[{\citenamefont{Haldane}(1995)}]{Haldane:prl95}
\bibinfo{author}{\bibfnamefont{F.~D.~M.} \bibnamefont{Haldane}},
  \bibinfo{journal}{Phys. Rev. Lett.} \textbf{\bibinfo{volume}{74}},
  \bibinfo{pages}{2090} (\bibinfo{year}{1995}).

\bibitem[{\citenamefont{Auerbach}(1998)}]{Auerbach:prl98}
\bibinfo{author}{\bibfnamefont{A.}~\bibnamefont{Auerbach}},
  \bibinfo{journal}{Phys. Rev. Lett.} \textbf{\bibinfo{volume}{80}},
  \bibinfo{pages}{817} (\bibinfo{year}{1998}).

\bibitem[{\citenamefont{Matveev}(1995)}]{MatveevQdot}
\bibinfo{author}{\bibfnamefont{K.~A.} \bibnamefont{Matveev}},
  \bibinfo{journal}{Phys. Rev. B} \textbf{\bibinfo{volume}{51}},
  \bibinfo{pages}{1743} (\bibinfo{year}{1995}).

\bibitem[{\citenamefont{Z\"ulicke et~al.}(1998)\citenamefont{Z\"ulicke,
  MacDonald, and Johnson}}]{Zulicke:prb98}
\bibinfo{author}{\bibfnamefont{U.}~\bibnamefont{Z\"ulicke {\it et al.}}},
  \bibinfo{journal}{Phys. Rev. B}
  \textbf{\bibinfo{volume}{58}}, \bibinfo{pages}{13778} (\bibinfo{year}{1998}).

\bibitem[{\citenamefont{Chklovskii et~al.}(1992)\citenamefont{Chklovskii,
  Shklovskii, and Glazman}}]{Chklovskii:prb92}
\bibinfo{author}{\bibfnamefont{D.~B.} \bibnamefont{Chklovskii {\it et al.}}},
  \bibinfo{journal}{Phys. Rev. B}
  \textbf{\bibinfo{volume}{46}}, \bibinfo{pages}{4026} (\bibinfo{year}{1992}).

\end{thebibliography}

\end{document}